\documentclass[reprint,amsmath,amssymb,aps]{revtex4}
\usepackage{graphicx}% Include figure files
\usepackage{dcolumn}% Align table columns on decimal point
\usepackage{bm}% bold math
\usepackage[english]{babel}
\begin{document}

\title{On the Velocity Tensors}
\author{E. Kapu\'scik}
\affiliation{Alfred Meissner Graduate School of Dental Engineering, Ustro\'n, Poland}
\email{edward.kapuscik@ifj.edu.pl}
\author{T. Lanczewski}
\affiliation{Institute of Nuclear Physics PAN, Krak\'ow, Poland}
\email{tomasz.lanczewski@ifj.edu.pl}

\begin{abstract}
\textbf{Abstract:} A new object, called the velocity tensor, is introduced. It allows to formulate a generally covariant mechanics. Some properties of the velocity
tensor are derived.
\end{abstract}

\maketitle

\section{Introduction}

In classical mechanics \cite{1} the velocity $\mathbf{v}\left(
t\right)$ of a material point is defined as
\begin{equation}
\label{1}\mathbf{v}\left(t\right) =\frac{d\mathbf{x}\left(
t\right)}{dt},
\end{equation}
where $\mathbf{x}\left(t\right)$ is the trajectory function of the
moving point and $t$ is the time coordinate in the chosen inertial reference
frame. In special relativity \cite{2} the notion of the three-dimensional
velocity $\mathbf{v}\left(t\right)$ is generalized to the notion
of the four-velocity defined as
\begin{equation}
\label{2}u^\mu\left(\tau\right)=\frac{dx^\mu\left(\tau\right)}{d\tau},
\end{equation}
where the space--time position of the material point is given by four
functions $x^\mu \left( \tau \right) (\mu=0,1,2,3; x^0 =ct)$ parameterized
by the so-called proper time
\begin{equation}
\label{3}d\tau =dt\sqrt{1-\frac{\mathbf{v}^2\left(t\right)}{c^2}}.
\end{equation}
Due to its dependence on velocity of the moving material point, the notion
of the proper time $\tau $ is different for each material point and for
non-uniform motions the proper time is not a uniformly changing function of
the coordinate time $t$. Moreover, for non-uniform motions the proper time
coincides with the coordinate time in continuously changing inertial
reference frames (the momentarily rest frames). Only for uniformly moving
material points the proper time coincides with the coordinate time in one
reference frame (the rest frame of the moving material point). In addition,
for many particle systems the trajectories of particles are parametrized by
different proper times and it is almost impossible to describe the
interaction between particles without the notion of propagating fields.
Therefore relativistic mechanics cannot be so well developed as the
nonrelativistic mechanics is.

Fortunately, there exists another way of passing from Galilean--Newtonian
mechanics to the relativistic one \cite{3} which is not based on Eq. (\ref{2}).
Indeed, it is easy to see that rewriting Eq. (\ref{1}) in the form
\begin{equation}
\label{4}d\mathbf{x}\left(t\right)-\mathbf{v}\left(
t\right)dt=0
\end{equation}
we can immediately generalize it to a relativistic (as a matter of fact,
generally) covariant form
\begin{equation}
\label{5}V_\nu ^\mu \left( x\right) dx^\nu =0,
\end{equation}
where a new mixed tensor field $V_\nu ^\mu \left( x\right) $ is introduced.
We shall name this tensor as {\it the velocity tensor}.

It is clear that for nontrivial velocity tensors ($V_\nu ^\mu \left(
x\right) \neq \delta _\nu ^\mu $) Eq. (\ref{5}) define some
submanifolds of the considered space--time. We shall require from the velocity
tensors that these submanifolds should always be one dimensional what means
that Eq. (\ref{5}) must determine some curves interpreted as
trajectories of the moving material points.

Form (\ref{5}) has the obvious advantage over (\ref{1}) and (\ref{2}),
that it does not use any evolution parameter and therefore it may be applied
to systems with arbitrary number of material points by generalizing (\ref{5})
to the set of relations
\begin{equation}
\label{7}V_{a,\nu }^\mu \left( x_a\right) dx_a^\nu =0,
\end{equation}
where the index $a$ labels different material points.

At each space--time event the velocity tensors (different for different
material points) fix the infinitesimal directions in which any material
point located at that event may move. In addition, {\it forms (\ref{5})
and (\ref{7}) are invariant under arbitrary changes of space--time
coordinates.} Therefore, they may be used to formulate a generally covariant
scheme for classical mechanics.

The aim of the present paper is to describe some interesting properties of
the velocity tensors. We shall also provide the explicit construction of the
general form of such tensors.

It is clear that velocity tensors are related to the kinematical part of
mechanics. We shall also touch the dynamical aspect of mechanics.

\section{General Properties of the Velocity Tensors}

Equation (\ref{5}), in $n$-dimensional space--time, is an eigen equation for
the $n\times n$-dimensional matrix $V$ (defined by the velocity tensor) for
the eigenvalue $0$, while the infinitesimal displacements $dx^\mu $ in any
motion are the eigenvectors of the velocity tensors belonging to this
eigenvalue.

Writing the characteristic equation for the general eigenvalue problem
\begin{equation}
\label{8}V_\nu ^\mu \left( x\right) dx^\nu =\lambda dx^\mu
\end{equation}
we get the equation for the possible eigenvalues $\lambda $
\begin{equation}
\label{9}\sum_{j=0}^n\left( -\lambda \right) ^{n-j}Tr_jV\left( x\right) =0,
\end{equation}
where $Tr_jV\left(x\right)$ denotes the sums of diagonal minors of order $j$
of the matrix $V\left(x\right).$ Obviously, $Tr_1V\left(x\right)$
coincides with the ordinary trace of $V\left(x\right)$ and $Tr_nV\left(
x\right)$ is the determinant of $V\left(x\right)$. For shortness, we also
use the convention
\begin{equation}
\label{10}Tr_0V\left( x\right) =1
\end{equation}
for any matrix $V\left( x\right) $.

Due to physical reason we must require that there should be only one
eigenvalue equal to $0$. This means that there should be a unique
eigenvector for any velocity tensor which fixes the infinitesimal
displacements in any motion. The characteristic equation (\ref{9}) must be
therefore of the form
\begin{equation}
\label{11}\lambda ^n=0,
\end{equation}
from which we get the following conditions for any velocity tensor:
\begin{equation}
\label{13}Tr_jV(x)=0
\end{equation}
for all $j>0.$

Conditions (\ref{13}) are generally covariant requirements because all
the $Tr_jV,$ being the coefficients in characteristic equation (\ref{9}%
), are invariant under arbitrary similarity matrix transformations and it is
well known that for mixed tensors, treated as matrices, the general
coordinate transformations locally become the similarity transformations

\begin{equation}
\label{14}V\left( x\right) \rightarrow V^{\prime }\left( x^{\prime }\right)
=S\left( x\right) V\left( x\right) S^{-1}\left( x\right) ,
\end{equation}
where the matrix elements of $S\left( x\right) $ are given by
\begin{equation}
\label{15}S_\nu ^\mu \left( x\right) =\frac{\partial x^{\prime\mu}\left(
x\right) }{\partial x^\nu }
\end{equation}
for arbitrary changes of space--time coordinates $x^\mu \rightarrow x^{
\prime\mu }\left( x\right) $ .

Conditions (\ref{13}) impose $n$ restrictions for the $n^2$ matrix elements
of the velocity tensors. Further restrictions come from the requirement
that, in each reference frame, from (\ref{5}) it should follow that
\begin{equation}
\label{16}dx^k=v^k\left( t\right)dt,
\end{equation}
where $k=1,...,(n-1)$ and $v^k\left( t\right) $ are the components of the
standard velocity. This gives us additional $n-1$ restrictions for the
matrix elements of the velocity tensor. Finally, we shall require that in $n$%
-dimensional spacetimes the motions in all $n-k$ subspaces should be
described exactly as they were described in the $\left( n-k\right) $-dimensional
spacetimes. This means that restricting the motions to subspaces
the form of the velocity tensor should reduce to the already established
forms of the velocity tensors in the corresponding lower dimensional subspaces. We shall
refer to this requirement as to the reduction principle. It is easy to count
that such a requirement gives additional $2^{n-1}-2$ conditions for the
matrix elements of any velocity tensor. Altogether we are left with
$n^2-n-\left( n-1\right) -\left( 2^{n-1}-2\right) =\left( n-1\right)
^2-\left( 2^{n-1}-2\right) $ free parameters of any velocity tensor. These
free parameters should represent components of some $\left( n-1\right)$-dimensional
vector which will guarantee the covariance of the velocity
tensor under space rotations because this is the only simple geometrical
interpretation of the remaining constants in the velocity tensors. In this
way, we arrive at the equation
\begin{equation}
\label{17}\left( n-1\right) ^2-\left( 2^{n-1}-2\right) =n-1.
\end{equation}
It is surprising that this equation has solution only for $n=2,3$ and $4$.
This means that our construction can be performed only in two, three and
four-dimensional spacetimes, correspondingly.

\section{General Construction of the Velocity Tensors}

We shall now present a simple method of the construction of all possible
velocity tensors.

Let us consider spacetimes for which the passage between inertial reference
frames is described by the linear change of coordinates

\begin{equation}
\label{17}x^\mu \rightarrow x^{\prime\mu }=L_\nu ^\mu \left(
\mathbf{u}\right) x^\nu ,
\end{equation}
where $\mu ,\nu =0,1,2,3$ and $\mathbf{u}$ denote the relative
velocity of the two inertial reference frames. From (\ref{17}) and the
tensor character of the velocity tensor we get the transformation law for it
(written in the matrix form)
\begin{equation}
\label{18}V\rightarrow V^{\prime }=L\left( \mathbf{u}\right)
VL^{-1}\left( \mathbf{u}\right) =L\left( \mathbf{u}\right)
VL\left( -\mathbf{u}\right) .
\end{equation}

It is clear that we should look for velocity tensors which are functions of the
ordinary velocity of motion. Our basic assumption consists in the
requirement that the functional forms of the velocity tensor are the same in
each reference frame. This means that
\begin{equation}
\label{19}V^{\prime }\left( \mathbf{v}^{\prime }\right) =V\left(
\mathbf{v}^{\prime }\right)
\end{equation}
because only under such condition in each reference frame we can fulfill
conditions (\ref{16}). In this way, transformation law (\ref{18}) becomes
to be a system of functional equations for the matrix elements of the matrix
$V$ of the following form:
\begin{equation}
\label{20}V\left( \mathbf{v}^{\prime }\right) =L\left(
\mathbf{u}\right) V\left( \mathbf{v}\right) L\left( -
\mathbf{u}\right) ,
\end{equation}
where
\begin{equation}
\label{21}v^{\prime k}=\frac{L_0^k\left( \mathbf{u}\right) +%
\sum_j L_j^k\left( \mathbf{u}\right)
v^j}{L_0^0\left( \mathbf{u}\right) + \sum_j
L_j^0\left( \mathbf{u}\right) v^j}.
\end{equation}

Taking into account that the particle at rest in the unprimed reference
frame moves with the velocity $-\mathbf{u}$ in the primed frame we
can rewrite these functional equations in the explicit form:
\begin{equation}
\label{22}V\left( \frac{-u^kL_0^0\left( \mathbf{u}\right) +\sum_j
L_j^k\left( \mathbf{u}\right) v^j}{
L_0^0\left( \mathbf{u}\right) +\sum_j
L_j^0\left( \mathbf{u}\right) v^j}\right) =L\left( \mathbf{u}
\right) V\left( \mathbf{v}\right) L\left( -\mathbf{u}\right).
\end{equation}

The solutions of these equations are obtained by the standard method. We
first put $v^k=0$, then change the signs of $u^k$ and finally rename $
\mathbf{u}$ into $\mathbf{v}$. As a result we get
\begin{equation}
\label{23}V\left( \mathbf{v}\right) =L\left( -\mathbf{v}%
\right) VL\left( \mathbf{v}\right) ,
\end{equation}
where on the right-hand side the matrix $V$ has constant matrix elements
equal to the elements of $V\left( 0\right) .$ The constant matrix elements
of $V$ should be determined by the additional requirements the velocity
tensors have to satisfy.

For all dimensions the first column of the velocity tensor $V$ consists of
null elements. This follows from the fact that for particles at rest the
eigenvector in (\ref{5}) is of the form
\begin{equation}
\label{24}\left(
\begin{array}{c}
dt \\
0 \\
0 \\
0
\end{array}
\right) .
\end{equation}

Such eigenvector will satisfy Eq. (\ref{5}) only if $V_0^\mu =0.$

\section{Examples}

\subsection{Two-Dimensional Space--Time}

For $n=2$ from conditions (\ref{13}) it follows that

\begin{equation}
\label{25}V=\left(
\begin{array}{cc}
0 & V_1^0 \\
0 & 0
\end{array}
\right) ,
\end{equation}
where $V_1^0$ is an arbitrary nonzero number. Since Eq.(\ref{5}) is
homogeneous, this constant can be taken as $1.$

For Galilean space--time
\begin{equation}
\label{26}L\left( u\right) =\left(
\begin{array}{cc}
1 & 0 \\
-u & 1
\end{array}
\right)
\end{equation}
and from (\ref{23}) we get
\begin{equation}
\label{27}V\left( v\right) =\left(
\begin{array}{cc}
-v & 1 \\
-v^2 & v
\end{array}
\right) .
\end{equation}

For Lorentz space--time
\begin{equation}
\label{28}L\left( u\right) =\frac 1{\sqrt{1-\frac{v^2}{c^2}}}\left(
\begin{array}{cc}
1 & -\frac u{c^2} \\
-u & 1
\end{array}
\right)
\end{equation}
and from (\ref{23}) we get
\begin{equation}
\label{29}V\left( v\right) =\frac 1{1-\frac{v^2}{c^2}}\left(
\begin{array}{cc}
-v & 1 \\
-v^2 & v
\end{array}
\right) .
\end{equation}

\subsection{Higher-Dimensional Spacetimes}

From the reduction principle and from the form of the velocity tensor in the
two-dimensional space--time we immediately get that in all higher dimensional
spacetimes the only non-zero components are the $V_k^0$. Therefore the final
form of the velocity tensors is
\begin{equation}
\label{30}V_\nu ^\mu \left( \mathbf{v}\left( t\right) \right)
=L_0^\mu \left( -\mathbf{v}\left( t\right) \right)
\sum_k V_k^0L_\nu ^k\left( \mathbf{v}\left(
t\right) \right) ,
\end{equation}
where $\left( V_1^0,V_2^0,....,V_n^0\right) $ are components of a $\left(
n-1\right) $-dimensional vector under rotations in the subspace $\left(
x^1,x^2,....,x^{n-1}\right) .$ Using this form of $V$ and the explicit forms of
the Galilean and Lorentz transformations we easily can get the velocity
tensor both for the Galilean and Lorentz spacetimes of any dimension.

\section{Dynamics}

Since our kinematical part of classical mechanics is generally covariant, it
is necessary to determine such form of dynamical equations which also will
be generally covariant. For this purpose we shall remind that the only
generally covariant differential relation which may be reduced to the famous
Newton relation
\begin{equation}
\label{27}\frac{d\mathbf{p}\left( t\right) }{dt}=\mathbf{F}%
\left( t\right)
\end{equation}
is of the form
\begin{equation}
\label{28}\nabla _\mu \pi ^{\mu \nu }\left( x\right) =F^\nu \left( x\right)
,
\end{equation}
where $\pi ^{\mu \nu }\left( x\right) $ is some tensorial density and $F^\nu
\left( x\right) $ is a vector density, while $\nabla _\mu $ denotes a
corresponding covariant derivative.

Assuming that $\pi ^{\mu \nu }\left( x\right) ,$ like the velocity tensors,
is a function of the ordinary velocity, we easily can construct the explicit
form of this quantity. This leads, exactly as for the velocity tensors, to
the following form of the dynamical tensor:
\begin{equation}
\label{29}\pi ^{\mu \nu }\left( \mathbf{v}\left( t\right) \right)
=L_\alpha ^\mu \left( -\mathbf{v}\left( t\right) \right) \pi
^{\alpha \beta }L_\beta ^\nu \left( -\mathbf{v}\left( t\right)
\right) ,
\end{equation}
where all $\pi ^{\alpha \beta }$ are constants. Since, in contradiction
to the velocity tensor, the dynamical tensor $\pi ^{\mu \nu }\left(
\mathbf{v}\right) $ need not to satisfy any additional conditions, we
have here to do with $n^2$ arbitrary constants which describe the inertial
properties of the considered particles. We may, however, diminish the number
of arbitrary constants by requiring the symmetry of $\pi ^{\mu \nu }\left(
\mathbf{v}\right) $ and then only one parameter, the mass of the
particle, describes its inertial property. In this case $\pi^{\mu \nu}$
simply is the energy-momentum tensor of the material point. Since the
$\pi ^{\mu \nu }\left(\mathbf{v}\left( t\right) \right) $ depends
only on the time coordinate, it is clear that Eq. (\ref{28}) reduces to
(\ref{27}).

\section{Conclusions}

We have introduced a new mechanical object called the velocity tensor and
explicitly constructed the velocity tensors in
spacetimes of any dimension. We hope that the notion of the velocity tensor
will shed more light on the possible dynamics in general relativity. It also
may be useful for relativistic many-body systems.

\end{document}